\documentstyle[12pt]{article}
\def\TL{\hfil$\displaystyle{##}$}
\def\TR{$\displaystyle{{}##}$\hfil}

\def\comment#1{}
\def\fixit#1{}

\def\overleftrightarrow#1{\vbox{\ialign{##\crcr
     $\leftrightarrow$\crcr\noalign{\kern-0pt\nointerlineskip}
     $\hfil\displaystyle{#1}\hfil$\crcr}}}

\def\lsim{\mathrel{\mathstrut\smash{\ooalign{\raise2.5pt\hbox{$<$}\cr\lower2.5pt\hbox{$\sim$}}}}}
\def\gsim{\mathrel{\mathstrut\smash{\ooalign{\raise2.5pt\hbox{$>$}\cr\lower2.5pt\hbox{$\sim$}}}}}

\def\sqr#1#2{{\vcenter{\vbox{\hrule height.#2pt
         \hbox{\vrule width.#2pt height#1pt \kern#1pt
            \vrule width.#2pt}
         \hrule height.#2pt}}}}

\def\href#1#2{#2}  

\def\lbldef#1#2{\expandafter\gdef\csname #1\endcsname {#2}}
\def\eqn#1#2{\lbldef{#1}{(\ref{#1})}%
\begin{equation} #2 \label{#1} \end{equation}}
\def\eqalign#1{\vcenter{\openup1\jot
    \halign{\strut\span\TL & \span\TR\cr #1 \cr
   }}}

\def\comment#1{  \begin{raggedright}{\tt [#1]}\end{raggedright}}
\def\fixit#1{}

\def\comment#1{  \begin{raggedright}{\tt [#1]}\end{raggedright}}
\def\fixit#1{}
\def\bm{{\bar{m}}}
\def\bn{{\bar{n}}}
\def\adg{a^{\dagger}}
\def\bC{\bar{C}}
\def\half{{ 1 \over 2 } }
\begin{document}
\baselineskip=15.5pt
\pagestyle{plain}
\setcounter{page}{1}


\begin{titlepage}

\begin{flushright}
PUPT-1980 \\  
hep-th/0104095
\end{flushright} 
\vfil

\begin{center}
{\Large Reverse Engineering ADHM Construction from Non-Commutative Instantons}
\end{center}
\vspace{1cm}
\begin{center}
{\large  Mukund Rangamani\footnote{rmukund@princeton.edu}}
\end{center}
\begin{center}   
Joseph Henry Laboratories, Princeton University, Princeton,
NJ 08544.
\end{center}  
\vfil

\begin{center}
{\large Abstract}
\end{center}  
\noindent
We study the non-commutative instanton solution proposed in 
hep-th/0009142 and obtain the spectrum of small oscillations.
The spectrum thus obtained is in exact agreement with the spectrum of 
stringy excitations in a configuration of point like D0 branes sitting on top 
of D4-branes with a uniform magnetic field turned on in the world-volume of the
D4-branes in the Seiberg-Witten decoupling limit. This provides
further evidence for the solution of hep-th/0009142 and also enables us recover
the ADHM data from the 0-4 string spectrum. 
Generalizations to higher co-dimension solitons are also discussed. 
\vfil
\vfil
\vspace{6cm}
\begin{flushleft}
April 2001.
\end{flushleft}

\end{titlepage}
\newpage

\section{Introduction}

\par
Non-commutative field theories have been the source of 
many interesting new physical insights. One of the many 
fascinating developments in this area has been the discovery 
of non-trivial solutions to the classical equations of motion
\cite{gms}. A striking feature is that the non-commutative deformation
permits solitonic solutions in theories which have no such in the 
commutative version. The study of non-trivial solutions in 
non-commutative field theories was initiated in \cite{gms}, 
and has been dealt by many authors in differing 
contexts \cite{a1}-\cite{an};  
see \cite{neklec, harvlec} for reviews on the subject 
and an extensive list of references. 

We shall in the present work be interested mainly in the 
properties of co-dimension four solitons {\it i.e.}, instantons 
in non-commutative gauge theories.
In the pioneering work of Nekrasov and Schwarz \cite{ns}, 
a simple deformation of the classic ADHM construction 
was shown to lead to the non-commutative instanton. An 
interesting feature of this instanton is the resolution of 
the small instanton singularity in the moduli space. 
A proof of this was presented in \cite{sw}, where 
from the sigma model description of the ADHM 
construction \cite{witsigma}, it was argued that the 
moduli space of self-dual non-commutative instantons 
depends only on the anti self-dual part of the non-commutativity 
parameter, $\Theta$. This implies that for self-dual 
$\Theta$ the moduli space has a small instanton singularity.
The fact may be understood by noting the absence of a supersymmetric 
bound state between a localized $Dp$ brane and a $D(p+4)$ brane
in the presence of a NS-NS two-form field, unless the $B$-field
is self-dual. 

In the present work we will provide a more 
direct evidence for the above chain of 
ideas. Our starting point will be the explicit construction of
self-dual instantons with self-dual non-commutativity \cite{agms}
\footnote{For recent work on non-commutative instantons see 
\cite{nikita,f,sch} }. 
We study the spectrum of small fluctuations about the 
instanton solution and reproduce the spectrum that one would 
obtain from a conformal field theory analysis for open strings 
in the Seiberg-Witten decoupling limit \cite{sw}. This would in 
principle amount to ``re-deriving'' the ADHM construction 
from the given explicit solution to the Yang-Mills equations of 
motion. It bears mentioning that this characterization of the 
instanton is a close cousin of the matrix theory description of 
the $Dp$-$D(p+4)$ bound state which was studied in \cite{gl}.
The solution of \cite{agms} is particularly interesting for it is 
easily generalized to constructing higher co-dimension solutions and 
one can carry out an analogous exercise in these cases too. 

Knowledge of the fluctuation spectrum for generic values of 
non-commutativity, where the solution is unstable, can be used 
to study the issue of tachyon condensation in the system.
In \cite{agms} the study of tachyon potential was carried out for the 
co-dimension 2 soliton. The tachyon potential 
in the $D0$-$D4$ story was studied in 
\cite{justin} from the string field theory perspective. 
One can use the present analysis to compute the exact tachyon 
potential as in the case of the co-dimension two soliton.
Given the generalization of the solution  to higher co-dimension
solitons it would provide us with an handle towards understanding
the physics governing the newly found supersymmetic bound states 
of $D0$-branes with $D6$-branes and $D8$-branes \cite{witten, cr,b1,b2,b3}.

The organization of the paper is as follows. 
In section 2 we review the construction of \cite{agms} and set forth our 
notation. We present the spectrum of fluctuations in 
section 3 and discuss the implications thereof. 
In section 4 higher co-dimension  solitons are discussed. 
Some of the algebra is relegated to the appendices to make for  
a more coherent discussion. 

\section{The instanton in non-commutative Yang-Mills} 
\par
In this section we review the basics of non-commutative gauge theory. 
In section 
2.1 we write down the Lagrangian of non-commutative gauge theory coupled to 
adjoint scalar fields in the operator language following 
\cite{gmst} and 
present the equations of motion. In the subsequent subsection we review the 
instanton solution of \cite{agms}. 
For most part of the discussion we shall focus on solutions with 
only the gauge field excited and the scalars shall be set to 
their vacuum value.

\subsection{Non-commutative Yang-Mills: A Review}
\par
We would like to consider $4+1$ dimensional non-commutative Yang-Mills (NCYM)
(we shall mainly focus on the $U(1)$ case)
theory with 5 adjoint scalars, to mimic the bosonic field content of 
the low-energy effective world-volume theory of D4-branes with a constant 
background magnetic field. The classical action in temporal gauge is 
given as 

\eqn{action}{ 
S = -{1 \over 4 g_{YM}^2} \int d^5 x \left( F_{\mu \nu} * F^{\mu \nu}
+ 2 \sum_{ i = 1}^5 D_{\mu} \phi^i * D^{\mu} \phi^i + \sum_{i=1}^{5} 
[\phi^i, \phi^j]_*[\phi^i,\phi^j]_* \right)
}

\noindent
We are on a non-commutative ${\bf R}^4$ with a non-commutativity
parameter $\Theta$ 
given by the block-diagonal form; $\Theta^{\mu \nu} = 
(\theta_1 \epsilon, \theta_2 \epsilon)$, (where $\epsilon$ is the
anti-symmetric $2 \times 2$ matrix with $\epsilon^{12} = 1$). 
We introduce complex coordinates $z^m$, obeying the commutation relation 
\eqn{defcood}{
[z^m, \bar{z}^{\bn}] = i \Theta^{m \bn}.
}

\noindent
We can exploit the relation between the algebra of functions on non-commutative
${\bf R}^4$ and the algebra of operators in the Hilbert space of a particle in 
2-spatial dimensions by defining ladder operators {\it c.f.} \cite{gmst} , 

\eqn{ladderops}{\eqalign{
a_{\bm} =  -i & \; \Theta^{-1}_{\bm n}  \; z^n; \;\;\;\;\;\; \adg_m = 
i \; \Theta_{m \bn}^{-1} \; \bar{z}^{\bn}, \cr
&[\adg_m , a_{\bn} ] = -i \; \Theta^{-1}_{m \bn}.
}}

\noindent
To recast the NCYM action \action in the operator language we parameterize the 
gauge field in terms of a operator in the Hilbert space $C_m$ and use the 
fact that translations can be implemented by taking commutators with respect to
the ladder operators. To wit, 
\eqn{defns}{\eqalign{
C_m &= -i A_m + a_m^{\dagger}; \;\;\;\;\;\; C_{\bm} = i A_{\bm} +
a_{\bm}, \cr
F_{m \bn} &= i [C_m , C_{\bn} ] - \Theta_{m \bn}^{-1}, \cr 
 D_m \phi &= - [C_m , \phi] ; \;\;\;\;\;\; D_{\bm} \phi = [C_{\bm},
\phi].
}}

\noindent
Finally we arrive at the action
\eqn{opact}{\eqalign{
S = -{ 4 \pi^2{\rm Pf}(\Theta )\over 4 g_{YM}^2} & \int dt \; {\rm Tr} 
\huge{\{} -\partial_t
\bC_{\bm}
\partial_t C_m - \sum_{i = 1}^{5} { 1 \over 2} \partial_t \phi^{i} 
\partial_t \phi^{i} \cr
& - 4 \left( i [C_m, \bC_{\bn}] -\Theta^{-1}_{m \bn} \right)^2
+ 8 \left(i [C_m, C_n]\right) \left( i [\bC_{\bm}, \bC_{\bn}]\right)  \cr
& +  \sum_{i=1}^{5} [C_m, \phi^i] [\phi^i,\bC_{\bm}] 
+ { 1 \over 4} \sum_{i,j=1}^{5} [\phi^i,\phi^j][\phi^i,\phi^j]  
\huge{\}}.
}}

\noindent
The equations of motion resulting from the variation of the action \opact\ are 
given as   
\eqn{eom}{\eqalign{
& \partial^2_t C_m =  [C_n, [C_m, \bC_{\bn} ]] +
   \sum_{i=1}^5 [\phi^i,[C_m, \phi^i]] \cr
& \partial^2_t \phi^i = [C_m, [\phi^i, \bC_{\bm}]] +
[\bC_{\bm},[\phi^i, C_m]]
+ \sum_{j=1}^{5} [\phi^j, [\phi^i, \phi^j]].
\cr
}}
\noindent
In addition to the equations of motion we also need to impose the
Gauss Law constraint 
to pick out the physical states and this reads,
\eqn{glaw}{
[\bC_{\bm}, \partial_t C_m] + [C_m, \partial_t \bC_{\bm}] +
\sum_{i=1}^5
[\phi^i, \partial_t \phi^j] = 0.
} 

\subsection{The instanton}
A static solution to the equations \eom\ was found in \cite{agms}.
This solution carries 
a single unit of $1^{st}$ Pontrjagin charge and is given as 
\eqn{instanton}{
C_m = T^{\dagger} \adg_m T; \;\;\;\;\;\; \phi_i = 0.
}
\noindent
Here $T$ is an operator obeying,
\eqn{trelns}{
T T^{\dagger}  = 1, \;\;\;\; T^{\dagger} T = 1 - P_0,  \;\;\;\; 
P_0 T^{\dagger} = TP_0 = 0. 
}
\noindent
$P_0$ is the  projection operator onto the ground state $\mid 0,0 \rangle$, 
of the two-particle system, {\it i.e.}, it projects onto 
lowest radial wavefunction in both complex directions.

The field strength for this configuration evaluates to 
$ F_{m \bn} = -\Theta_{m \bn}^{-1} P_0$ (note that the 
vacuum with zero field strength in this notation is $C_m = \adg_m$),
and this implies that the first Pontrjagin class of the solution is $\pm 1$ 
depending on whether $\Theta_{\mu \nu}$ is self-dual or anti-self-dual. 
Evaluating the energy of the solution one finds 
$ S = 2 \pi^2 g_{YM}^2 \; \sqrt{{\rm det} \Theta}
\; \Theta^{-1}_{m \bn} \; \left(\Theta^{-1} \right)^{m \bn}$, 
which saturates the BPS 
bound iff $\Theta$ is self-dual or anti-self dual. We shall consider the case 
of instanton number one solution and hence we shall be interested in 
particular at the situation with self-dual $\Theta$. 
Based on the energetics and the charge of the solution it was conjectured in 
\cite{agms} that the solution \instanton\ corresponds to 
an anti-zero brane localized at a point on the four-brane
\footnote{We shall persist in referring to the system as that of a $D0$-brane 
localized on a $D4$-brane}. 
Evidence for this was offered in \cite{f} by directly solving the 
ADHM equations.

To complete the characterization of the solution we shall present an 
explicit representation of the operator $T$. In the standard number basis 
of states for a two-dimensional harmonic oscillator,
\eqn{basisam}{
\mid n_1,n_2 \rangle = { \left( \adg_1 \right)^{n_1} 
\left( \adg_2 \right)^{n_2}  \over
\sqrt{\left(n_1 \right)! \left( n_2\right)!}} \mid 0,0 \rangle,
}
\noindent
we can define an integer ordering of states as follows:
\eqn{intord}{
\mid k = n_1 + {(n_1 +n_2 )(n_1 +n_2 +1) \over 2} \rangle \;  
\equiv \; \mid n_1,n_2 
\rangle.
}
$T$ can then be represented as a shift operator with respect to this ordering, 
its matrix elements being given as 
\eqn{tmat}{
\langle k \mid T \mid l \rangle = \delta_{k,l-1}.
}
\noindent

\section{Classical Fluctuation analysis}
The solution given in Eq \instanton\ is supposed to be a co-dimension 4 
soliton in a $4+1$ dimensional NCYM theory. In terms of a brane picture 
it should correspond to a localized $D0$-brane on the world-volume of a 
$D4$-brane with constant $B$-field. 
This configuration is generically not supersymmetric for arbitrary 
values of $\Theta$, but for self-dual $\Theta$ we obtain a supersymmetric 
configuration \cite{ns, sw}. This is the special point in the moduli space of 
non-commutative instantons where the moduli space is the same as the 
commutative case, and the solution we have corresponds to this small 
instanton point. 

The analysis of the string spectrum in this background was done in
\cite{sw} and we refer to them for the basic results. It was found that 
the low lying modes in the decoupling limit are the standard 
massless modes of the $0-0$ and $4-4$ strings 
along with additional modes from the $0-4$ with masses proportional 
to the non-commutativity parameter. Typically, in usual configurations of 
d-branes, taking the low energy limit leads to keeping only the massless modes 
of the strings. In the presence of the $B$-field however, the presence of an 
additional dimensionful parameter which is being scaled appropriately to 
preserve the non-commutativity, leads to additional set of modes which 
have masses related to the non-commutativity scale. 
In particular one has low lying modes with masses given by 

\eqn{cftmass}{\eqalign{
\half E_{1,2(-)} 
&= \pm \left({1 \over \theta_1} - {1 \over \theta_2}\right) \cr
\half E_{1+} & = {3 \over \theta_1} + {1 \over \theta_2} \cr
\half E_{2+} & = {1 \over \theta_1 } +{ 3 \over \theta_2} \cr
\half E_i &= { 1 \over \theta_1} + { 1 \over \theta_2} \;\;\;\;\; i = 5
,\cdots, 9.
}}

\noindent
The rest of the spectrum can be worked out by acting on the ground state with 
the oscillators. 

The spectrum we present here is valid for generic values of 
the non-commutativity 
parameter. In particular note that there is a tachyonic mode 
(from $E_{1,2(-)}$ 
depending on the relative strengths of $\theta_1$ and $\theta_2$) which 
becomes massless along with another massive mode as $\Theta$ 
approaches the self-dual 
value. This is the indicator of restoration of supersymmetry as the solution 
becomes BPS for this special point \cite{ns, sw}. 

As mentioned, the system under consideration can be understood from a 
matrix theory 
point of view. In matrix theory one writes down classical brane configurations 
in terms of non-commuting matrices which satisfy the equations of 
motion.  One builds a configuration of $D4$-branes by 
having say, $[X_1, X_2] = i \theta_1$ and $[X_3,X_4] = i \theta_2$.
This can be extended to include $D0$-branes by letting the matrices 
have a block diagonal form with the upper block being given by the 
above form to make up the $D4$-branes and the lower block being a 
diagonal matrix with the eigenvalues parametrising the position of the 
$D0$-branes. This configuration was studied by Lifschytz \cite{gl},
and the fluctuation spectrum was found to be as follows;
\eqn{matmass}{\eqalign{
\half E_{1(\pm)} & = \left({2n_1 + 1 \over \theta_1} + { 2 n_2 + 1 \over \theta_2} \right)
\pm {2 \over \theta_1}  \cr
\half E_{2(\pm)} & = \left({2n_1 + 1 \over \theta_1} + { 2 n_2 + 1 \over \theta_2} \right)
\pm {2 \over \theta_2}  \cr
\half E_{i} & = \left({2n_1 + 1 \over \theta_1} + { 2 n_2 + 1 \over \theta_2} \right)
}}
\noindent
It bears mentioning that the matrix theory manipulations are closely related to
the non-commutative gauge theory ones \cite{gmst,seiberg}, for,
rewriting the non-commutative 
gauge theory in the operator language is akin to working with the dissolved 
$D0$-branes in the world-volume of the $D4$-branes.  
We shall exploit this correspondence to parameterize the fluctuations in a
canonical fashion. 

\subsection{Scalar fluctuations}
In the solution given by \instanton\ the scalars are unexcited, which makes
it easy for us to study the spectrum arising from their fluctuations. 
We shall detail the calculation for the case with the scalars unexcited 
and indicate the modifications arising when the scalars are given a 
non-trivial vev. As is clear from the geometric picture the scalars we 
are considering correspond to the directions transverse to the $D4$-brane.
This means that we shall be presently considering the situation when the 
$D0$-brane is sitting right on top of the $D4$-brane. In generic 
situation we could move the $D0$-brane away giving additional contributions 
to the mass coming from the string having to stretch the extra distance. 
  
Without loss of generality we can focus on the case of a single scalar.
We choose to parameterize the fluctuation of the scalar 
$\delta \phi$ as ({\it c.f.}, \cite{agms} )
\eqn{scadecomp}{
\delta \phi = \chi + \psi + \bar{\psi} + T^{\dagger}\gamma T
}
\noindent
where, 
\eqn{sd}{\eqalign{
  \chi &
= P_0 \delta \phi P_0 , \;\;\;\;\; \psi = P_0 \delta \phi
(1-P_0), \cr
  T^{\dagger} \gamma T &
= (1-P_0) \delta \phi ( 1- P_0).
}}
\noindent
The logic here is to separate the fluctuation into components such that 
the separation between the various string sectors is made manifest. 
From a matrix theory point of view this corresponds to the fact that the 
$D0$ and the $D4$ brane solutions are in different blocks along the 
diagonal and the off-diagonal piece is related to the $ND$ strings of 
the $0-4$ sector. As already mentioned, the operator language used to write 
solutions in 
NCYM is closed related to the matrix formulation \cite{seiberg}.
Denoting the complete Hilbert space by ${ \cal{H}}$, let 
${\cal{H}}_0$ be the subspace to which $P_0$ projects, and 
${\cal{H}}_{\perp}$ be the orthogonal component {\it i.e.}, 
${ \cal { H} } = { \cal {H}}_0 \cup {\cal{H}}_{\perp}$.
Then the analogy with the matrix description is made manifest
by associating the ${\cal{H}}_0$ with the 
$0$-brane block and the ${\cal{H}}_{\perp}$ with the $4$-brane 
block. So the $0$-$4$ strings arise from the modes which are project 
from one side into ${\cal{H}}_0$ and from the other into ${\cal{H}}_{\perp}$. 

 Given this parameterization we can evaluate the scalar potential to
quadratic order about the solution \instanton. The contributions to the 
potential at this order are going to come from  
the term $ Tr [ C_m , \phi][\phi , \bC_{\bm}]$. Plugging in the form 
suggested by \scadecomp , we find
\eqn{scalarpot}{ 
{\rm Tr} [ C_m , \delta \phi][\delta \phi , \bC_{\bm}] = 
{\rm Tr} \left( \left( C_{(0)m} \bC_{(0)\bm} + \bC_{(0)\bm} C_{(0)m} \right)
 \bar{\psi } \psi + [\adg_m , \gamma][ \gamma, a_{\bm}] \right).
}

The field $\chi$ does not appear in the scalar potential at quadratic 
order. Hence a massless scalar from the $0+1$ dimensional point
of view. These are the fluctuation modes of the scalars on the
$D0$-brane in the directions transverse to the $D4-$brane.
This gels well with the intuition gained from comparison to matrix theory. 
The modes represented
by $\gamma$ have the right  potential to be the transverse scalars on the 
world-volume of the $D4-$brane. In particular the commutators with the 
creation-annihilation operators is exactly what is necessary to 
covariantize the derivatives.
This leaves us with the fields $\psi$ which are the $0-4$ scalars.
Their mass is given by the eigenvalues of the operator 
$\left( C_{(0)m} \bC_{(0)\bm} + \bC_{(0)\bm} C_{(0)m} \right)$. 
This is simple to evaluate in the integer ordered basis \intord\
for the Hilbert space. Expanding the field $\psi$; 

\eqn{psiexpan}{
\psi = \sum_{k=0}^{\infty} \psi_k \mid 0 \rangle \langle k + 1 \mid,
}

we can write the relevant term in \scalarpot\ as 

\eqn{psiev}{\eqalign{
 & \sum_{k,l=0}^{\infty} \psi_k \bar{\psi}_l \langle k + 1  \mid 
T^{\dagger} \adg_m T T^{\dagger} a_{\bm} T + T^{\dagger} a_{\bm} T 
T^{\dagger} \adg_m T \mid l + 1 \rangle \cr
& = \sum_{n_i = 0}^{\infty} 
\mid \psi_{\{ n_1 n_2 \}} \mid^2 \langle n_1, n_2 \mid \adg_m a_{\bm} + a_{\bm}
\adg_m
\mid n_1,n_2 \rangle \cr
& = \sum_{n_i = 0}^{\infty}
 \left( {2n_1 + 1 \over \theta_1} + { 2n_2 + 1 \over \theta_2} 
\right)\mid \psi_{\{j,m\}} \mid^2 
}}

\noindent
In the above series of manipulations we have used the fact that 
$T$ acts as a lowering operator in the basis \intord. 
As a result of this we find that the summation extends over the 
whole Hilbert space of states and one can conveniently switch over 
from the basis \intord\ to the standard number basis 
for purposes of evaluating the matrix elements. 
From \psiev\ we find that there is a low-lying mode
$\psi_{\{0,0\}}$, of mass ${1 \over \theta_1} + { 1 \over \theta_2}$ and 
over that we have a whole tower of massive modes.
This matches perfectly with the CFT analysis of \cite{sw}, given in 
Eq \cftmass\,  
and the  matrix theory calculation of \cite{gl}, Eq \matmass. 
Note that in this analysis the Gauss law constraint \glaw\ plays no 
role as the background solution \instanton\ has the scalar field 
unexcited.

\subsection{Gauge field fluctuations}
The gauge field fluctuations can be analyzed in a fashion
analogous to the scalar fluctuations. As before we decompose
the fluctuations as 
\eqn{gff}{\eqalign{
C_m &= C_{(0)m } + \delta C_m \cr
\delta C_m &= A_m + W_m + \bar{Q}_m + T^{\dagger} D_m T,
}}
\noindent
with, 
\eqn{gfexp}{\eqalign{
& A_m = P_0 \delta C_m P_0 , \;\;\;\;\;\; W_m = P_0 \delta C_m (1-P_0), \cr
&\bar{Q}_m = (1-P_0) \delta C_m P_0 , \;\;\;\;\;\; T^{\dagger} D_m T 
 = (1 - P_0) \delta C_m (1-P_0).
}}
The potential for the gauge field fluctuations comes from the term 
${ 1 \over 2} \left( i [C_m, \bC_{\bn}] - \Theta^{-1}_{m \bn}
\right)^2$.
Substituting the above form of the fluctuations we find the
contribution to the potential to be 
\eqn{gfpot}{
L(W,Q) 
+ {1 \over 2} \left([\adg_m, \bar{D}] + [D, a_{\bm}] \right)^2.
}
\noindent
The explicit form for $L(W,Q)$ is given in the appendix. 
One can find appropriate linear combinations of the fields 
$W_{1,2}$ and $Q_{\bar{1},\bar{2}}$, labeled $U,V,X,Y$  
in terms of which the Lagrangian $L(W,Q)$  is easily diagonalized. 

As before with the case of the scalar fluctuations, from the absence  
of quadratic terms for the field $A_m$,
we are led to conclude that these are the 
massless modes corresponding to the motion of the $D0$-brane
along the world-volume of the $D4$-brane.
The field $D$ is the gauge field on the world-volume of the $D4-$brane.
To analyze the spectrum of the off-diagonal modes, $W$ and $Q$ we expand
them in the harmonic oscillator basis; taking appropriate 
linear combinations (see Appendix for details) we find the spectrum,

\eqn{mass}{\eqalign{
\half E_{\{n_1,n_2\}}(U)  & = {(2n_1 + 1)\over \theta_1} + 
{(2n_2 +1) \over \theta_2}+ {2 \over \theta_1}\cr
\half E_{\{n_1,n_2\}}(V) & =  {(2n_1 + 1 ) \over \theta_1} + 
{(2n_2 +1)\over \theta_2}+{2\over \theta_2} \cr
\half E_{\{n_1,n_2\}}(X) & = {(2n_1+1)\over \theta_1} + 
{(2n_2+1)\over \theta_2} - { 2\over \theta_1} \cr
\half E_{\{n_1,n_2\}}(Y) & = {(2n_1+1)\over \theta_1} + 
{(2n_2+1)\over \theta_2} - { 2\over \theta_2} 
}}

\noindent
This is indeed isomorphic to the spectrum given in \matmass.
Apart from these modes we also have the $D0$-brane gauge field, 
which arises from the component $A_0$, of the gauge potential in the 
$D4$-brane theory. 
In all one has the complete spectrum to be the standard $0$-$0$ 
and $4$-$4$ string spectrum and the spectrum of the $0$-$4$ strings 
as given in \psiev\ and \mass.

\subsection{Relating to the ADHM construction}
We have obtained the spectrum of fluctuations for a single 
point-like $D0$-brane in a $4+1$ dimensional
$U(1)$ non-commutative gauge theory. 
Generalizing the above to multi-instanton configurations is 
simply achieved by writing down the solution \instanton\ with 
the operator $T$ replaced by $T^k$, $k$ being the number of 
instantons. The spectrum in this case works out just the same, 
from the gauge field fluctuations we get the $0$-$0$ sector 
gauge fields which are in the adjoint of $U(k)$, and also 
scalars corresponding to motion of the $D0$-brane along the 
four-brane which too are in the adjoint of $U(k)$. 
On the contrary the $0$-$4$ strings are charged in the 
fundamental representation of $U(k)$. If we had considered 
a non-abelian generalization by having $N$ $D4$-branes, 
then we would have the $0$-$4$ strings charged in the fundamental of 
$U(N)$. This is pretty much all the information we need 
to reconstruct the ADHM data by just following the chain of logic 
in section 5 of \cite{sw}. 

The low energy effective theory of the system is 
the quantum mechanics of the $0$-$4$ strings and the 
$0$-$0$ strings. The relevant modes are the  
adjoint scalars (the scalars which correspond to motion in 
directions transverse to the $D4$-brane; the scalars in the 
directions tangential to the $D4$-brane
are Goldstone modes of the translational symmetry and decouple from the 
low energy dynamics), the gauge 
field from the $0$-$0$ sector and the two low-lying modes 
$Q_{\bar{1}, \bar{2}\{ 0,0\}}$ with masses 
$\pm \left({ 1 \over \theta_1} - {1 \over \theta_2}\right)$
from the $0$-$4$ sector. The adjoint scalars are denoted as 
$x$ and $y$ and the fundamental scalars are $p$ and $q$ 
respectively. The tachyonic mode in fact serves to determine 
the strength of the FI coupling in this theory. The classical 
potential in this framework is given as 

\eqn{adhmpot}{ 
{ \rm Tr} \left\{ \left( [x,x^{\dagger}] + [y, y^{\dagger}] + q a^{\dagger}
- p^{\dagger} p - \zeta \right)^2 + \mid \; [x,y] + q p \; \mid^2 \right\}. 
}

\noindent
$\zeta$ is the FI coupling, determined from the existence of a tachyonic 
mode in the spectrum of the $0$-$4$ strings at generic values of $\Theta$. 
For self-dual $\Theta$ this vanishes. Hence although the 
locus $x = y = p = q = 0$ is not a solution to the minimum of the 
potential at generic $\Theta$, it is indeed present when $\Theta$ is 
self-dual as then $\zeta$ vanishes. 
The ADHM equations are just the standard equations of motion for 
the quantum mechanics theory as is well known. 
Thus having verified that the spectrum of fluctuations about the 
instanton background \instanton\ is indeed as predicted by 
a conformal field theory analysis, we have managed to recover all the
ingredients essential to reverse engineer the ADHM construction. 

\section{Generalization to higher dimensions}
Recently, it was shown by Witten \cite{witten}
that one can in the presence of non-commutativity have supersymmetric
bound states of $D0$-branes with $D6$-branes and $D8$-branes
\footnote{In the absence of a B-field the $D0$-$D6$ is non-supersymmetric,
while the $D0$-$D8$ admits a susy bound state. In the presence of 
non-commutativity there is a second supersymmetric branch of the 
$D0$-brane with a $D8$-brane.}. These systems were also discussed in
\cite{cr,b1,b2,b3}. 

One can easily generalize the solution of \cite{agms} to obtain these 
higher co-dimension solitons. It is clear that in the case of a
co-dimension $2n$ soliton ($n= 3,4$), one needs to find the 
analog of the shift operator $T$ in a $n$-particle Hilbert space. 
In the $n$-particle Hilbert space one may similarly introduce 
an integer ordering of the basis states and introduce a shift operator 
with respect to that ordering. As we saw in our analysis the 
explicit form for the operator was not quite essential in determining 
the spectrum of small fluctuations. 

The scalar mass spectrum is given by the eigenvalues of the 
operator \\
$\left( C_{(0)m} \bC_{(0)\bm} + \bC_{(0)\bm} C_{(0)m} \right)$. 
This masses work out to be  $2 \left({( 2 n_1 +1 ) \over \theta_1} + 
{( 2 n_2 +1 ) \over \theta_2} +{( 2 n_3 +1 ) \over \theta_3} \right)$ for 
the $D0$-$D6$ case. To obtain the gauge field fluctuations one would have to 
do a little more work, but the end result is simple. In case 
of the $D0$-$D6$ configuration we obtain, 

\eqn{zsmass}{\eqalign{
\half
E_{1}^{\pm} &= {( 2 n_1 +1 ) \over \theta_1} + {( 2 n_2 +1 ) \over \theta_2} + 
{( 2 n_3 +1 ) \over \theta_3} \pm {2 \over \theta_1} \cr
\half
E_{2}^{\pm} &= {( 2 n_1 +1 ) \over \theta_1} + {( 2 n_2 +1 ) \over \theta_2} + 
{( 2 n_3 +1 ) \over \theta_3} \pm {2 \over \theta_2} \cr
\half
E_{3}^{\pm} &= {( 2 n_1 +1 ) \over \theta_1} + {( 2 n_2 +1 ) \over \theta_2} + 
{( 2 n_3 +1 ) \over \theta_3} \pm {2 \over \theta_3} 
}}

\noindent 
and for the $D0$-$D8$ we have

\eqn{zemass}{\eqalign{
\half
E_{1}^{\pm} &= {( 2 n_1 +1 ) \over \theta_1} + {( 2 n_2 +1 ) \over \theta_2} + 
{( 2 n_3 +1 ) \over \theta_3}
+{( 2 n_4 +1 ) \over \theta_4} \pm {2 \over \theta_1} \cr
\half
E_{2}^{\pm} &= {( 2 n_1 +1 ) \over \theta_1} + {( 2 n_2 +1 ) \over \theta_2} + 
{( 2 n_3 +1 ) \over \theta_3} +{( 2 n_4 +1 ) \over \theta_4}
\pm {2 \over \theta_2} \cr
\half
E_{3}^{\pm} &= {( 2 n_1 +1 ) \over \theta_1} + {( 2 n_2 +1 ) \over \theta_2} + 
{( 2 n_3 +1 ) \over \theta_3}+  {( 2 n_4 +1 ) \over \theta_4}
 \pm {2 \over \theta_3} \cr
\half
E_{4}^{\pm} &= {( 2 n_1 +1 ) \over \theta_1} + {( 2 n_2 +1 ) \over \theta_2} + 
{( 2 n_3 +1 ) \over \theta_3} + {( 2 n_4 +1 ) \over \theta_4}
\pm {2 \over \theta_4} 
}}

Yet again from the knowledge of the fluctuation spectrum we 
can write down the low energy effective theory governing the dynamics 
of the bound state. It would be given by the quantum mechanics of 
the $0$-$0$ strings interacting with the $0$-$6$ or $0$-$8$ strings,
depending on the case of interest.  

\section*{Acknowledgments}
I would like to thank A.~Bergman, O.~Ganor, R.~Gopakumar, S.~Gukov,
A.~Hashimoto, C.~Herzog, V.~Hubeny, 
S.~Minwalla, S.~Murthy and especially 
E.~Witten for extremely fruitful discussions. 
In addition I would like to thank Caltech for hospitality during the concluding
stages of the project. This work was supported in part by NSF grant PHY-980248.

\section*{Appendix A }
In this appendix we present the details of the calculation pertaining to the 
gauge field fluctautions. 

The potential for the gauge field fluctuations comes from the terms of 
the kind ${ 1 \over 2} \left( i [C_m, \bC_{\bn}] - \Theta^{-1}_{m \bn}
\right)^2$. Substituting the form of the fluctuations as in \gfexp\ we get
(modulo an overall factor of ${ {\rm Pf}(\Theta)  \over g_{YM}^2}$)

\eqn{gfpota}{
{1 \over 2} \left([\adg_m, \bar{D}] + [D, a_{\bm}] \right)^2
+ L(W,Q)
}

\noindent
Expanding the off-diagonal fields in the mode expansion as in the 
scalar case \psiexpan

\eqn{QWexp}{\eqalign{
& W_{m} = \sum_{k=0}^{\infty} W_{m (k)} \mid 0 \rangle \langle k +1
\mid, \cr
& Q_{\bm} = \sum_{k=0}^{\infty} Q_{\bm (k)} \mid 0 \rangle \langle k
+1 \mid.
}}

\noindent
$L(W,Q)$ can be written as 

\eqn{WQfint}{\eqalign{
&\sum_{k,l= 0}^{\infty} 
W_{1(k)} \bar{W}_{{\bar{1}} (l)} \; \langle k \mid a_{{\bar{1}}} \adg_1
+ a_{\bar{2}} \adg_2 + \adg_2 a_{\bar{2}} \mid l \rangle
+ Q_{{\bar{1}} (k)} \bar{Q}_{1 (l)} \; \langle k \mid \adg_1 a_{\bar{1}}
+\adg_2 a_{\bar{2}} + a_{\bar{2}} \adg_2 \mid l \rangle \cr
&+  W_{2(k)} \bar{W}_{{\bar{2}} (l)} \; \langle k \mid a_{{\bar{2}}} \adg_2
+ a_{\bar{1}} \adg_1 + \adg_1 a_{\bar{1}} \mid l \rangle
+ Q_{{\bar{2}} (k)} \bar{Q}_{2 (l)} \; \langle k \mid \adg_2 a_{\bar{2}}
+ \adg_1 a_{\bar{1}} + a_{\bar{1}} \adg_1 \mid l \rangle \cr
& - \left( W_{1(k)} \bar{Q}_{1(l)} \; \langle k \mid a_{{\bar{1}}} a_{{\bar{1}}} \mid l
\rangle + W_{1(k)} \bar{Q}_{2(l)} \; \langle k \mid a_{{\bar{2}}} a_{{\bar{2}}} \mid l
\rangle + c.c \right)  \cr
& - \left( W_{2(k)} \bar{Q}_{2(l)} \; \langle k \mid a_{{\bar{2}}} a_{{\bar{2}}} \mid l
\rangle + W_{2(k)} \bar{Q}_{1(l)} \; \langle k \mid a_{{\bar{1}}} a_{{\bar{1}}} \mid l
\rangle + c.c \right)  \cr
& - \left( W_{1(k)} \bar{W}_{{\bar{2}}(l)} \; \langle k \mid a_{{\bar{1}}} \adg_{2} \mid l
\rangle + Q_{{\bar{1}}(k)} \bar{Q}_{2(l)} \; \langle k \mid a_{{\bar{2}}} \adg_{1} \mid l
\rangle + c.c \right) \cr
& + i \Theta^{-1}_{m \bn} \left(\sum_{k=0}^{\infty} W_{m (k)} \bar{W}_{\bn (k)}
- Q_{\bn (k)} \bar{Q}_{m (k)}\right) .
}}

\noindent
We shall in the following set $\Theta$ to unity to avoid notational clutter.
As before we only have to evaluate the matrix elements. We find

\eqn{WQfluc}{\eqalign{
L (W,Q) & = \sum_{n_i=0}^{\infty} 
( n_1 + 2n_2 + 3) \; W_{1 \{n_1,n_2\} } \bar{W}_{\bar{1} \{n_1,n_2\} }  
+ ( 2n_1 + n_2 +3  ) \;
W_{2 \{n_1, n_2 \} } \bar{W}_{\bar{2} \{ n_1,n_2\} }  \cr
& +(n_1 + 2n_2 ) \; Q_{\bar{1} \{n_1,n_2\} } \bar{Q}_{1 \{ n_1,n_2\} } 
+ (n_2 + 2n_1) \;Q_{\bar{2} \{n_1,n_2\} } \bar{Q}_{2 \{ n_1,n_2\} } \cr
& - \left( \sqrt{n_2(n_1 +1)} \; W_{1 \{n_1, n_2\} }
 \bar{W}_{\bar{2} \{ n_1 +1 , n_2 -1 \}}
+ \sqrt{n_1 (n_2 +1 ) } \;
Q_{{\bar{1}} \{n_1,n_2\} } \bar{Q}_{1 \{ n_1 -1,n_2 +1  \} }
+ c.c. \right) \cr
& - \sqrt{(n_1+1)(n_2+1)} \left( W_{1 \{n_1,n_2\} } \bar{Q}_{2 \{ n_1 +1,n_2 +1
\} }
+ W_{2 \{n_1,n_2\} } \bar{Q}_{1 \{ n_1+1,n_2+1 \} } + c.c\right) \cr
& - \sqrt{(n_1+1)(n_1+2)} \left( W_{1 \{n_1,n_2\} } \bar{Q}_{1 \{ n_1+2,n_2
\} } + c.c \right) \cr
& - \sqrt{(n_2+1)(n_2+2)} \left( W_{2 \{n_1,n_2\} } \bar{Q}_{2 \{ n_1,n_2 +2
\} } + c.c \right) 
}}

It is useful to 
introduce linear combinations of the fields $W$ and $Q$, 

\eqn{diagonal}{\eqalign{
U_{\{n_1,n_2\}}  &= \sqrt{n_1 + 2n_2 +3 } \; W_{1 \{n_1, n_2\}}
- \sqrt{{(n_1 +1)(n_1 +2) \over (n_1 + 2n_2 +3)}}
\; \bar{Q}_{1 \{ n_1 +2, n_2\}} \cr
& - \sqrt{{(n_1 + 1 )(n_2 + 1 )\over (n_1 + 2n_2 +3) }}
\; \bar{Q}_{2 \{n_1 + 1 ,n_2 +1\}} - 
\sqrt{{n_2 (n_1 +1 )  \over (n_1 + 2n_2 +3) }} \; 
\bar{W}_{\bar{2} \{n_1 +1 , n_2 -1 \}} \cr
V_{\{n_1, n_2\}} & = \sqrt{{2 (n_1 + n_2 +2) \over (n_1 + 2n_2 +4 ) }}
(\sqrt{(n_1 + n_2 +3)} \; W_{2 \{n_1, n_2 \}} - \sqrt{(n_2 + 1 )(n_2+2)}
\; \bar{Q}_{2 \{n_1,n_2+2\}} \cr
& \;\;\;\;- \sqrt{(n_1 + 1 )(n_2+1)} \; \bar{Q}_{1 \{n_1 +1, n_2 +1 \}} ) \cr
X_{\{n_1,n_2\}} & = \sqrt{{2 (n_1 +n_2) \over (n_1 + n_2 +1)}}
\left( \sqrt{n_2 +1} \; Q_{\bar{1} \{n_1,n_2\}} - \sqrt{n_1}\; Q_{\{n_1 -1, 
n_2 + 1\}} \right) \cr
Y_{\{n_1,n_2\}} &= Q_{\bar{2}\{n_1, n_2\}}
}}

\noindent
leading to a simple form for the Lagrangian;
\eqn{maslag}{\eqalign{
L(W,Q) & = (2n_1 + 2n_2 +4) \; \mid U_{\{n_1,n_2\} }\mid^2
+ (2n_1 + 2n_2 +4) \; \mid V_{\{n_1,n_2\} }\mid^2 \cr
& \;\;\;+ (2n_1 + 2n_2 ) \; \mid X_{\{n_1,n_2\}} \mid^2
+ (2n_1 + 2n_2 ) \; \mid Y_{\{n_1,n_2\} }\mid^2
}}

\noindent
It is sufficiently simple to reintroduce the appropriate powers of 
$\theta_1, \theta_2$ into the expressions for the masses and we end up with 
the result given in \mass. 

Howeve, there are a couple of subtleties that need to be addressed.
From \WQfluc\ we see that the modes 
$Q_{\bar{1},\bar{2} \{0,0\}}$ are already diagonal and have 
masses $\pm \left({ 1 \over \theta_1} - { 1 \over \theta_2} \right) $.
The modes $Q_{\bar{1},\bar{2}
\{1,0\}}$ and $Q_{\bar{1},\bar{2} \{0,1\}}$ on the other hand only mix 
amongst themselves, but they too have the right masses to 
fit into the general scheme given in \mass.
One other issue to worry about is that of the linear 
combinations orthogonal to the ones introduced in \diagonal. If these modes 
were physical then our correspondence would be destroyed by the 
presence of a large number of massless modes. Fortunately for us 
this is not the case, these modes can be shown to be pure gauge and 
hence are unphysical. This is easily seen by write out the 
Gauss law constraint \glaw\ in terms of the modes introduced 
and finding that the aforementioned modes have vanishing time 
derivatives.




\begin{thebibliography}{99}

\bibitem{gms}
R.~Gopakumar, S.~Minwalla and A.~Strominger,
``Noncommutative solitons,''
JHEP{\bf 0005}, 020 (2000)
[hep-th/0003160].

\bibitem{ns}
N.~Nekrasov and A.~Schwarz,
``Instantons on noncommutative R**4 and (2,0) superconformal six  dimensional theory,''
Commun.\ Math.\ Phys.\ {\bf 198}, 689 (1998)
[hep-th/9802068].

\bibitem{a1}
D.~J.~Gross and N.~A.~Nekrasov,
``Monopoles and strings in noncommutative gauge theory,''
JHEP{\bf 0007}, 034 (2000)
[hep-th/0005204].

\bibitem{a2}
A.~P.~Polychronakos,
``Flux tube solutions in noncommutative gauge theories,''
Phys.\ Lett.\ B {\bf 495}, 407 (2000)
[hep-th/0007043].

\bibitem{a3}
D.~P.~Jatkar, G.~Mandal and S.~R.~Wadia,
``Nielsen-Olesen vortices in noncommutative Abelian Higgs model,''
JHEP{\bf 0009}, 018 (2000)
[hep-th/0007078].

\bibitem{Bak:2000ym}
D.~Bak and K.~Lee,
``Elongation of moving noncommutative solitons,''
Phys.\ Lett.\ B {\bf 495}, 231 (2000)
[hep-th/0007107].

\bibitem{a4}
A.~S.~Gorsky, Y.~M.~Makeenko and K.~G.~Selivanov,
``On noncommutative vacua and noncommutative solitons,''
Phys.\ Lett.\ B {\bf 492}, 344 (2000)
[hep-th/0007247].

\bibitem{a5}
C.~Zhou,
``Noncommutative scalar solitons at finite Theta,''
hep-th/0007255.

\bibitem{Lindstrom:2000kh}
U.~Lindstrom, M.~Rocek and R.~von Unge,
``Non-commutative soliton scattering,''
JHEP{\bf 0012}, 004 (2000)
[hep-th/0008108].

\bibitem{a6}
A.~Solovyov,
``On noncommutative solitons,''
Mod.\ Phys.\ Lett.\ A {\bf 15}, 2205 (2000)
[hep-th/0008199].

\bibitem{a7}
D.~Bak,
``Exact multi-vortex solutions in noncommutative Abelian-Higgs theory,''
Phys.\ Lett.\ B {\bf 495}, 251 (2000)
[hep-th/0008204]

\bibitem{agms} 
M.~Aganagic, R.~Gopakumar, S.~Minwalla and A.~Strominger,
``Unstable solitons in noncommutative gauge theory,''
hep-th/0009142.

\bibitem{a8}
J.~A.~Harvey, P.~Kraus and F.~Larsen,
``Exact noncommutative solitons,''
JHEP{\bf 0012}, 024 (2000)
[hep-th/0010060].

\bibitem{nikita}
N.~A.~Nekrasov,
``Noncommutative instantons revisited,''
hep-th/0010017.

\bibitem{f}
K.~Furuuchi,
``Dp-D(p+4) in noncommutative Yang-Mills,''
hep-th/0010119.

\bibitem{a9}
D.~J.~Gross and N.~A.~Nekrasov,
``Solitons in noncommutative gauge theory,''
hep-th/0010090.

\bibitem{c1}
M.~Hamanaka and S.~Terashima,
``On exact noncommutative BPS solitons,''
JHEP{\bf 0103}, 034 (2001)
[hep-th/0010221].

\bibitem{c2}
K.~Hashimoto,
``Fluxons and exact BPS solitons in non-commutative gauge theory,''
JHEP{\bf 0012}, 023 (2000)
[hep-th/0010251].

\bibitem{neklec}
N.~A.~Nekrasov,
``Trieste lectures on solitons in noncommutative gauge theories,''
hep-th/0011095.

\bibitem{c3}
D.~Bak, S.~U., K.~Lee and J.~Park,
``Noncommutative vortex solitons,''
hep-th/0011099.

\bibitem{c4}
B.~Durhuus, T.~Jonsson and R.~Nest,
``Noncommutative scalar solitons: Existence and nonexistence,''
Phys.\ Lett.\ B {\bf 500}, 320 (2001)
[hep-th/0011139].

\bibitem{neklec}
N.~A.~Nekrasov,
``Trieste lectures on solitons in noncommutative gauge theories,''
hep-th/0011095.

\bibitem{Li:2000ig}
M.~Li,
``Quantum corrections to noncommutative solitons,''
hep-th/0011170.

\bibitem{harvlec}
J.~A.~Harvey,
``Komaba lectures on noncommutative solitons and D-branes,''
hep-th/0102076.

\bibitem{Sahraoui:2000jq}
E.~M.~Sahraoui and E.~H.~Saidi,
``Solitons on compact and noncompact spaces in large noncommutativity,''
hep-th/0012259.

\bibitem{Martinec:2001hh}
E.~J.~Martinec and G.~Moore,
``Noncommutative solitons on orbifolds,''
hep-th/0101199.

\bibitem{Kiem:2001ny}
Y.~Kiem, C.~Kim and Y.~Kim,
``Noncommutative Q-balls,''
hep-th/0102160.

\bibitem{sch}
A.~Schwarz, 
``Noncommutative instantons: A new approach,''
hep-th/0102182.

\bibitem{Jackson:2001iy}
M.~G.~Jackson,
``The stability of noncommutative scalar solitons,''
hep-th/0103217.

\bibitem{Gopakumar:2001yw}
R.~Gopakumar, M.~Headrick and M.~Spradlin,
``On noncommutative multi-solitons,''
hep-th/0103256.

\bibitem{an}
L.~Hadasz, U.~Lindstrom, M.~Rocek and R.~v.~Unge,
``Noncommutative multisolitons: Moduli spaces, quantization, finite Theta  effects and stability,''
hep-th/0104017.

\bibitem{sw}
N.~Seiberg and E.~Witten,
``String theory and noncommutative geometry,''
JHEP{\bf 9909}, 032 (1999)
[hep-th/9908142].

\bibitem{witsigma}
E.~Witten,
``Sigma models and the ADHM construction of instantons,''
J.\ Geom.\ Phys.\ {\bf 15}, 215 (1995)
[hep-th/9410052].




\bibitem{gl}
G.~Lifschytz,
``Four-brane and six-brane interactions in M(atrix) theory,''
Nucl.\ Phys.\ B {\bf 520}, 105 (1998)
[hep-th/9612223].

\bibitem{seiberg}
N.~Seiberg,
``A note on background independence in noncommutative gauge theories,  matrix model and tachyon condensation,''
JHEP{\bf 0009}, 003 (2000)
[hep-th/0008013].

\bibitem{justin}
J.~R.~David,
``Tachyon condensation in the D0/D4 system,''
JHEP{\bf 0010}, 004 (2000)
[hep-th/0007235].

\bibitem{witten}
E.~Witten,
``BPS bound states of D0-D6 and D0-D8 systems in a B-field,''
hep-th/0012054.

\bibitem{cr}
M.~Mihailescu, I.~Y.~Park and T.~A.~Tran,
``D-branes as solitons of an N = 1, D = 10 non-commutative gauge theory,''
hep-th/0011079.

\bibitem{b1}
M.~Sato,
``BPS bound states of D6-branes and lower dimensional D-branes,''
hep-th/0101226.

\bibitem{b2}
K.~Ohta,
``Supersymmetric D-brane bound states with B-field and higher dimensional  instantons on noncommutative geometry,''
hep-th/0101082.

\bibitem{b3}
R.~Blumenhagen, V.~Braun and R.~Helling,
``Bound states of D(2p)-D0 systems and supersymmetric p-cycles,''
hep-th/0012157.

\bibitem{gmst}
R.~Gopakumar, S.~Minwalla and A.~Strominger,
``Symmetry restoration and tachyon condensation in open string theory,''
hep-th/0007226.

\end{thebibliography}
\end{document}